\documentclass[12pt]{article}

\usepackage[utf8]{inputenc}
\usepackage[margin=1in]{geometry}
\usepackage{amsmath,amssymb,amsthm}
\DeclareMathOperator*{\argmin}{arg\,min}
\usepackage{graphicx}
\usepackage{booktabs}
\usepackage{comment}
\usepackage{natbib}
\usepackage{hyperref}
\usepackage{caption}
\usepackage{subcaption}
\usepackage{float}
\usepackage{setspace}
\usepackage{threeparttable}

\title{\vspace{-2cm}Two-Step Regularized HARX to Measure Volatility Spillovers in Multi-Dimensional Systems\vspace{-0.5cm}}

\author{Mindy L. Mallory\footnote{Associate Professor and Clearing Corporation Foundation Chair in Food and Agricultural Marketing, Department of Agricultural Economics, Purdue University, West Lafayette, IN 47907. Email: mlmallor@purdue.edu}}

\date{January 2026}

\begin{document}

\maketitle

\begin{abstract}
We identify volatility spillovers across commodities, equities, and treasuries using a hybrid HARX-ElasticNet framework on daily realized volatility for six futures markets over 2002--2025. Our two step procedure estimates own-volatility dynamics via OLS to preserve persistence, then applies ElasticNet regularization to cross-market spillovers. The sparse network structure that emerges shows equity markets (ES, NQ) act as the primary volatility transmitters, while crude oil (CL) ends up being the largest receiver of cross-market shocks. Agricultural commodities stay isolated from the larger network. A simple univariate HAR model achieves equally performing point forecasts as our model, but our approach reveals network structure that univariate models cannot. Joint Impulse Response Functions trace how shocks propagate through the network. Our contribution is to demonstrate that hybrid estimation methods can identify meaningful spillover pathways while preserving forecast performance.
\end{abstract}



\newpage

\section{Introduction}

During the market chaos of March 2020 equity markets were crashing, crude oil prices were swinging wildly, and grain traders were watching their screens nervously. Soybean futures sometimes seem to catch volatility from crude oil and other 'outside' markets. This time, however, soybean futures remained fairly calm compared to what was going on in other markets.  One of the most interesting features of financial markets in modern times is the potential for cross-market volatility spillovers among markets that have merely a tangential relationship.  Understanding \textit{which} cross-market linkages actually matter (and which are just noise) is essential for risk management, portfolio construction, and systemic risk monitoring. The problem? Existing approaches face a fundamental tension. Unrestricted multivariate models overfit, but univariate models ignore cross-market dynamics entirely.

This paper asks: can we identify which cross-market volatility linkages matter?  We estimate a Heterogeneous Autoregressive (HAR) model using a hybrid two-step procedure:  first we estimate own-volatility dynamics via OLS to preserve realistic persistence of volatility through time (around 0.99), then we apply ElasticNet regularization to cross-market spillover terms to select economically meaningful spillovers while shrinking noise. The resulting sparse network reveals the structure of volatility transmission across six markets: soybeans, crude oil, the S\&P 500, the Nasdaq, and 5-year and 10-year treasury futures.  We then trace out how a volatility shock in one market propagates to others using Joint Impulse Response Functions, which allow us to shock multiple markets at once, and do not require onerous identification assumptions.  This matters because real-world events—a Fed announcement, a geopolitical crisis—hit several markets simultaneously and can amplify or cancel volatility spillovers depending on the direction of influence.

The volatility spillover literature has been dominated by GARCH-based methods since \citet{engle1982autoregressive} and \citet{bollerslev1986generalized} established the foundational framework.  Extensions to capture asymmetric responses to positive and negative shocks include the GJR-GARCH model of \citet{glosten1993relation}, which remains widely used for modeling the leverage effect in equity volatility. For multivariate settings, \citet{engle2002dynamic} introduced DCC-GARCH, which models time-varying correlations parsimoniously, while \citet{engle1995multivariate} developed the BEKK parameterization that allows explicit modeling of volatility transmission across assets.  These multivariate GARCH specifications remain the dominant way to study volatility spillovers—appearing in thousands of empirical applications across many market contexts.

A parallel literature has developed around the spillover index of \citet{diebold2009measuring} and spillover shares of \citet{diebold2012better}, which uses variance decompositions from vector autoregressions to measure connectedness across markets. This approach has an advantage.  It provides intuitive measures of directional spillovers (how much of asset A's forecast error variance is attributable to shocks from asset B) without imposing the parametric structure of GARCH.  \citet{barunik2018measuring} extended this framework to the frequency domain, decomposing spillovers into short-run and long-run components. 

For realized volatility forecasting specifically, \citet{corsi2009simple} developed the Heterogeneous Autoregressive (HAR) model, capturing the long-memory properties of volatility using a simple additive structure of daily, weekly, and monthly volatility components. The HAR model has become a workhorse for univariate volatility forecasting.  But it doesn't address cross-market spillovers. Extensions incorporating exogenous variables, like the HAR-X framework, have been developed by \citet{DEGIANNAKIS201728}, and \citet{audrino2016lassoing} demonstrate that LASSO regularization effectively identifies sparse coefficient structures implied by the HAR model. However, these extensions typically focus on forecasting rather than impulse response analysis.

A growing literature applies regularization methods to HAR and VAR models for spillover analysis.  \citet{demirer2018estimating} demonstrated the use of LASSO-VAR within the \citet{diebold2012better} spillover framework. With this approach they were able to estimate the high-dimensional network of 150 global banks. For commodity markets, \citet{yang2021volatility} apply LASSO-VAR to 25 commodity futures, finding that energy commodities serve as primary spillover transmitters.  \citet{ma2019forecasting} show that ElasticNet and LASSO combinations outperform the oil price volatility benchmark. \citet{audrino2016lassoing} provide theoretical and empirical support for applying LASSO to HAR models, showing that regularization effectively recovers sparse structures without sacrificing forecast accuracy. \citet{ding2021forecasting} find that multivariate LASSO models outperform univariate HAR at longer forecast horizons and introduce ordered LASSO to respect time series structure.

Our approach bridges these literatures by developing a hybrid HARX-ElasticNet framework that addresses two distinct econometric challenges in high-dimensional volatility modeling: capturing realistic persistence dynamics and identifying sparse cross-market spillover networks.  Standard unrestricted VAR estimation becomes infeasible when the number of variables approaches the sample size. With six markets and three HAR horizons per market, our model includes 108 potential parameters. Our two-step procedure first estimates univariate Heterogeneous Autoregressive (HAR) models via OLS for each market's own-volatility dynamics, preserving the characteristic high persistence documented in the realized volatility literature \citep{corsi2009simple}. We then apply ElasticNet regularization to the cross-market terms, which set 70--75 percent of spillover parameters to zero.  This sparsity is economically sensible—while a handful of central markets exhibit significant cross-linkages, most market pairs have negligible direct volatility transmission. By separating own-dynamics from cross-market selection, our approach combines the strengths of HAR models (realistic persistence) with regularization methods (sparse network identification), yielding impulse responses that reveal volatility spillover structure.  

We find three main results.  First, we identify a sparse spillover network where most cross-market transmission is zero. Only 7 of 90 cross-market coefficients (8\%) are nonzero.  Critically, equities (ES, NQ) and treasuries (ZF, ZN) exhibit \textit{zero} cross-market spillovers in both directions. The surviving spillovers are concentrated in commodity markets, where crude oil (CL) receives small spillovers from equities and treasuries, while soybeans (ZS) remain largely isolated.  

Second, the parsimonious HAR model and our hybrid HARX-ElasticNet achieve \textit{effectively identical} out-of-sample RMSE (0.0044).  This equivalence reflects the hybrid model's sparsity. For forecasting purposes, the hybrid model performs similarly to the univariate HAR. This demonstrates that cross-market information contributes negligibly to predictive accuracy while revealing economically meaningful network structure.  

Third, Joint Impulse Response Functions (JIRFs) illustrate how correlated volatilities can amplify and cross-propagate over time.  The JIRF methodology does not identify causal channels; rather, it traces the reduced-form dynamics of joint volatility movements implied by the estimated covariance structure.  Despite sparse coefficient estimates, JIRFs reveal substantial co-movement: when equity volatility rises, crude oil volatility tends to rise as well (from 0.04 to 0.09 over 20 days), reflecting the historical correlation between these markets.  The accumulating response patterns and widening confidence intervals highlight how correlated shocks can compound through the system, even when direct spillover coefficients are small. 

Our contribution is both methodological and empirical.  We extend the regularized HAR literature to a multi-asset-class setting spanning commodities, equities, and treasuries, and combine spillover network identification with Joint IRF analysis for policy-relevant insights into cross-market volatility transmission. To the best of our knowledge, this is the first application of the JIRF concept to a multivariate volatility modeling exercise. 

The rest of the paper proceeds as follows.  Section 2 describes our data and how we construct realized volatility using the Yang-Zhang estimator. Section 3 lays out the hybrid HARX-ElasticNet model along with the Joint Impulse Response methodology.  Section 4 presents results on forecast performance and spillover patterns.  Section 5 concludes.

\section{Data and Realized Volatility}

This section describes our data sources and the construction of realized volatility measures.  We use daily futures prices for six markets spanning commodities, equities, and treasuries, and compute volatility using the Yang-Zhang estimator,   which exploits overnight and intraday price movements for more efficient estimation than close-to-close returns alone.

\subsection{Data Sources}

We obtain daily futures data from Barchart OnDemand API for the period May 14, 2002 to January 31, 2025, yielding 5,699 observations after computing 30-day rolling Yang-Zhang volatility estimates. Our sample includes:

\begin{itemize}
    \item \textbf{Commodities}: Soybeans (ZS), Crude Oil (CL)
    \item \textbf{Equity Indices}: S\&P 500 (ES), Nasdaq-100 (NQ)
    \item \textbf{Treasury Futures}: 5-Year (ZF), 10-Year (ZN)
\end{itemize}

We use nearby continuous contracts with back-adjustment and roll 30 days before expiration to avoid delivery effects.  We select these six markets to span three major asset classes while keeping the model tractable for demonstration purposes, even though the regularization approach would afford a much larger model. Soybeans and crude oil represent agricultural and energy commodities, respectively—two sectors with distinct supply-side drivers but potential linkages through biofuel markets and broader macroeconomic conditions.  The S\&P 500 and Nasdaq-100 capture U.S. equity market volatility. The 5-year and 10-year treasury futures represent the intermediate portion of the yield curve, where monetary policy expectations and flight-to-quality flows are most evident.

\begin{figure}[H]
\centering
\includegraphics[width=\textwidth]{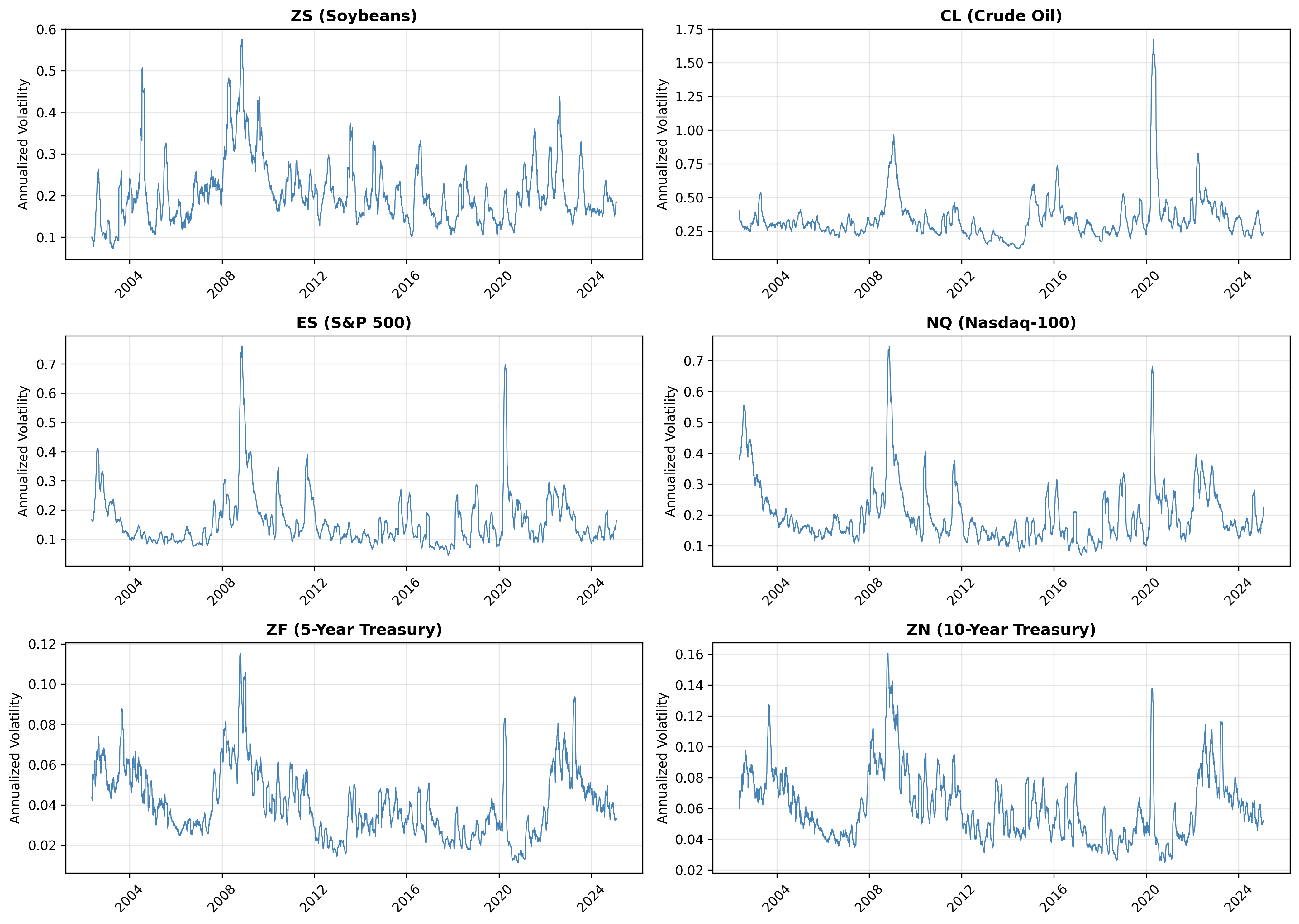}

\caption{Realized Volatility Time Series (Yang-Zhang Estimator). \small\textit{Notes: 30-day rolling Yang-Zhang realized volatility estimates for six futures markets. Sample period: May 2002 to January 2025 (5,699 observations). All volatilities are annualized.}}
\label{fig:rv_series}
\end{figure}

\subsection{Realized Volatility Estimation}

We compute realized volatility using the Yang-Zhang estimator \citep{yang2000drift}, which is considered the most efficient range-based volatility estimator.  Using a 30-day rolling window:
\begin{equation}
    RV_t^{YZ} = \sqrt{\sigma_o^2 + k \cdot \sigma_{oc}^2 + (1-k) \cdot \sigma_{RS}^2} \times \sqrt{252}
\end{equation}
where $\sigma_o^2$ is the overnight variance (variance of $\log(O_t/C_{t-1})$), $\sigma_{oc}^2$ is the open-to-close variance (variance of $\log(C_t/O_t)$), and $\sigma_{RS}^2$ is the rolling mean of the Rogers-Satchell \citep{rogers1991estimating} daily variance:
\begin{equation}
    \sigma_{RS,t}^2 = \log\frac{H_t}{C_t} \cdot \log\frac{H_t}{O_t} + \log\frac{L_t}{C_t} \cdot \log\frac{L_t}{O_t}
\end{equation}
Here $H_t$, $L_t$, $C_t$, $O_t$ denote daily high, low, close, and open prices, respectively.  The parameter $k = 0.34/(1.34 + (n + 1) / (n - 1))$ where $n=30$ is the rolling window size, chosen to minimize variance. The estimator is annualized by multiplying by $\sqrt{252}$.

The Yang-Zhang estimator is more efficient than Garman-Klass \citep{garman1980estimation} or Parkinson \citep{parkinson1980extreme} estimators because it is robust to opening jumps (overnight gaps) while still using all available daily price information.  \citet{yang2000drift} show it has the minimum variance among unbiased range-based estimators for assets that exhibit opening jumps—which is common in futures markets.

Table \ref{tab:summary_stats} presents summary statistics for realized volatility across assets.

\begin{table}[H]
\centering
\caption{Summary Statistics: Realized Volatility (Annualized)}
\label{tab:summary_stats}
\begin{tabular}{lrrrrrr}
\toprule
Asset & Mean & Std Dev & Min & Max & Skew & Kurt \\
\midrule
ZS & 0.207 & 0.077 & 0.072 & 0.575 & 1.47 & 2.92 \\
CL & 0.340 & 0.168 & 0.119 & 1.673 & 3.82 & 21.78 \\
ES & 0.161 & 0.093 & 0.044 & 0.761 & 2.83 & 11.54 \\
NQ & 0.201 & 0.098 & 0.068 & 0.746 & 2.14 & 6.28 \\
ZF & 0.041 & 0.017 & 0.011 & 0.115 & 1.00 & 1.37 \\
ZN & 0.061 & 0.022 & 0.025 & 0.161 & 1.16 & 1.71 \\
\bottomrule
\end{tabular}

\vspace{0.5em}
\parbox{\textwidth}{\small\textit{Notes: Sample period is May 14, 2002 to January 31, 2025 (5,699 observations). Realized volatility computed using Yang-Zhang estimator with 30-day rolling windows, annualized by $\sqrt{252}$.}}
\end{table}

Augmented Dickey-Fuller tests strongly reject the unit root null for all six series at the 5\% level, confirming stationarity (see Table \ref{tab:stationarity} in the Online Supplement).

Volatility is highest for crude oil (mean 34.0\%) and lowest for 5-year treasury futures (mean 4.1\%).  All series exhibit positive skewness and excess kurtosis, consistent with volatility clustering and occasional spikes during crisis periods.

\section{Methodology}

We employ a hybrid two-step estimation procedure that separates own-volatility dynamics from cross-market spillovers.  This approach preserves realistic volatility persistence while identifying sparse spillover networks.

\subsection{The HAR Model for Own-Volatility Dynamics}

Realized volatility exhibits two stylized facts that inform our modeling approach: high persistence (autocorrelations that decay slowly over many lags) and apparent long memory \citep{corsi2009simple}.  The Heterogeneous Autoregressive (HAR) model captures these properties parsimoniously by aggregating volatility at daily, weekly, and monthly horizons:
\begin{equation}
\label{eq:har}
RV_{i,t} = \alpha_i + \beta_i^{(d)} RV_{i,t-1} + \beta_i^{(w)} RV_{i,t-1:t-5}^{(w)} + \beta_i^{(m)} RV_{i,t-1:t-22}^{(m)} + u_{i,t}
\end{equation}
where $RV_{i,t-1:t-5}^{(w)} = \frac{1}{5}\sum_{j=1}^{5} RV_{i,t-j}$ and $RV_{i,t-1:t-22}^{(m)} = \frac{1}{22}\sum_{j=1}^{22} RV_{i,t-j}$ denote weekly and monthly RV averages.

The HAR specification yields persistence coefficients $\phi_i = \beta_i^{(d)} + \beta_i^{(w)} + \beta_i^{(m)}$ typically in the range 0.95--0.99 for commodity and equity futures, implying that volatility shocks decay slowly with half-lives of 14--70 trading days.  This high persistence is economically meaningful: it reflects the clustering of market uncertainty and the gradual resolution of information.  Preserving realistic persistence is essential for impulse response analysis—overly rapid decay would understate the duration of volatility transmission.

\subsection{The High-Dimensional Spillover Problem}

Measuring volatility spillovers across $K=6$ markets using HAR components requires estimating cross-market coefficients.  With three HAR components (daily, weekly, monthly) per market, the full specification for each market $i$ includes:
\begin{equation}
RV_{i,t} = \alpha_i + \sum_{j=1}^{K} \left[ \beta_{ij}^{(d)} RV_{j,t-1} + \beta_{ij}^{(w)} RV_{j,t-1:t-5}^{(w)} + \beta_{ij}^{(m)} RV_{j,t-1:t-22}^{(m)} \right] + \varepsilon_{i,t}
\end{equation}
where $\beta_{ij}$ captures the effect of market $j$'s lagged volatility on market $i$.  With $K=6$ markets and 3 components each, this specification contains 18 parameters per equation—or 108 total parameters across the system.

Regularization methods such as LASSO and ElasticNet address this dimensionality challenge by penalizing coefficient magnitudes, shrinking many to zero.  However, applying regularization uniformly to all coefficients creates a tension: the penalty that achieves appropriate sparsity in cross-market terms (setting 70--80\% to zero) simultaneously overshrinks own-lag coefficients, destroying the persistence structure documented above.

In pilot estimation, we found that standard ElasticNet applied to the full coefficient matrix reduced own-lag persistence from approximately 0.99 to 0.08--0.18.  Impulse responses from such models decay within 2--3 days—inconsistent with observed volatility dynamics and economically implausible for understanding volatility transmission.  This finding motivates our two step approach.

\subsection{Hybrid Two-Step Estimation Procedure}

We develop a two-step procedure that separates the estimation of own-volatility dynamics from cross-market spillover identification:

\paragraph{Step 1: HAR Estimation for Own-Lag Dynamics.}
For each market $i$, estimate the univariate HAR model in equation~\eqref{eq:har} by ordinary least squares.  This yields own-lag coefficients $\hat{\beta}_i^{(d)}$, $\hat{\beta}_i^{(w)}$, and $\hat{\beta}_i^{(m)}$ that preserve the characteristic high persistence of realized volatility.  Denote the fitted own-lag contribution as:
\begin{equation}
\widehat{RV}_{i,t}^{\text{own}} = \hat{\alpha}_i + \hat{\beta}_i^{(d)} RV_{i,t-1} + \hat{\beta}_i^{(w)} RV_{i,t-1:t-5}^{(w)} + \hat{\beta}_i^{(m)} RV_{i,t-1:t-22}^{(m)}
\end{equation}

\paragraph{Step 2: ElasticNet for Cross-Market Spillovers.}
Compute residuals $\tilde{u}_{i,t} = RV_{i,t} - \widehat{RV}_{i,t}^{\text{own}}$ and regress these on the lagged volatility of all \emph{other} markets:
\begin{equation}
\label{eq:step2}
\tilde{u}_{i,t} = \sum_{j \neq i} \left[ \gamma_{ij}^{(d)} RV_{j,t-1} + \gamma_{ij}^{(w)} RV_{j,t-1:t-5}^{(w)} + \gamma_{ij}^{(m)} RV_{j,t-1:t-22}^{(m)} \right] + \xi_{i,t}
\end{equation}
Estimate equation~\eqref{eq:step2} via ElasticNet:

\begin{equation}
\label{eq:elasticnet}
\begin{split}
\hat{\gamma} = \argmin_{\gamma} \Bigg\{ &\sum_{t} \left( \tilde{u}_{i,t} - \sum_{j \neq i} \left[ \gamma_{ij}^{(d)} RV_{j,t-1} + \gamma_{ij}^{(w)} RV_{j,t-1:t-5}^{(w)} + \gamma_{ij}^{(m)} RV_{j,t-1:t-22}^{(m)} \right] \right)^2 \\
&+ \lambda \left[ \alpha \|\gamma\|_1 + \frac{1-\alpha}{2} \|\gamma\|_2^2 \right] \Bigg\}
\end{split}
\end{equation}

where $\lambda > 0$ controls overall shrinkage and $\alpha \in (0,1)$ balances $L_1$ (sparsity) and $L_2$ (grouping) penalties.  We select $(\lambda, \alpha)$ via 5-fold time-series cross-validation, using non-overlapping sequential folds to respect temporal dependence.

\paragraph{Final Model.}
Combining both steps, the estimated model takes the form:

\begin{equation}
\label{eq:final}
\begin{split}
RV_{i,t} = &\underbrace{\hat{\alpha}_i + \hat{\beta}_i^{(d)} RV_{i,t-1} + \hat{\beta}_i^{(w)} RV_{i,t-1:t-5}^{(w)} + \hat{\beta}_i^{(m)} RV_{i,t-1:t-22}^{(m)}}_{\text{Own-lag dynamics (OLS, high persistence)}} \\
&+ \underbrace{\sum_{j \neq i} \left[ \hat{\gamma}_{ij}^{(d)} RV_{j,t-1} + \hat{\gamma}_{ij}^{(w)} RV_{j,t-1:t-5}^{(w)} + \hat{\gamma}_{ij}^{(m)} RV_{j,t-1:t-22}^{(m)} \right]}_{\text{Cross-market spillovers (ElasticNet, sparse)}} + \hat{\xi}_{i,t}
\end{split}
\end{equation}

\paragraph{Economic Rationale.}
The hybrid procedure separates two distinct economic phenomena that warrant different statistical treatments.  \textit{Own-volatility persistence} reflects market-specific information processing. News announcements, supply shocks, and idiosyncratic events generate volatility clustering that persists regardless of cross-market linkages. This component should exhibit high persistence, a property well-captured by HAR's multi-horizon aggregation and OLS estimation without shrinkage. \textit{Cross-market spillovers} reflect economic and informational linkages: common macroeconomic exposures, portfolio rebalancing, and contagion during stress episodes. Most market pairs have negligible direct linkages; only a sparse subset of connections transmit volatility materially.  ElasticNet regularization identifies this sparse network structure.  By fixing own-lag coefficients before estimating spillovers, we ensure that impulse response functions exhibit realistic dynamics: shocks to market $i$ generate persistent own-effects (reflecting HAR decay patterns) combined with sparse cross-transmission (reflecting network structure).

\subsection{HAR Benchmark}

Our primary benchmark is the univariate HAR model of \citet{corsi2009simple}.  For each asset $i$:
\begin{equation}
    RV_{i,t} = \beta_0 + \beta_1 RV_{i,t-1} + \beta_2 \overline{RV}^{(5)}_{i,t-1} + \beta_3 \overline{RV}^{(22)}_{i,t-1} + \varepsilon_{i,t}
\end{equation}

\subsection{Reduced Form Spillover Network }

The advantage of the hybrid HARX-ElasticNet framework is that the estimated coefficient matrix directly reveals the reduced form spillover network structure.  The $\gamma$ coefficients indicate which cross-market transmission channels are economically meaningful (those that survive regularization) versus which are noise that gets shrunk to zero.

\subsection{Joint Impulse Response Functions}

Traditional IRFs measure the response to a one-unit shock in a single variable. Joint Impulse Response Functions (JIRF), developed by \citet{wiesen2023joint}, generalize this framework to simultaneous shocks across multiple variables.  This extension is particularly valuable for studying financial market stress, where events such as Federal Reserve announcements or geopolitical crises typically affect multiple markets at once rather than one in isolation.  The JIRF methodology allows us to trace out the dynamic effects of such multi-market shocks while properly accounting for the joint structure of the initial impulse.

Let $\mathcal{S} \subset \{1, \ldots, K\}$ denote a subset of shocked variables (e.g., all equity indices).  The JIRF at horizon $h$ is defined as $JIRF_j^{\mathcal{S}}(h) = E[y_{j,t+h} | \text{shock to } \mathcal{S} \text{ at } t] - E[y_{j,t+h} | \text{no shock}]$.

\textbf{Joint Shock Construction.}   When shocking multiple variables simultaneously (e.g., both ES and NQ in an equity shock), we must specify how the shocks are correlated. Following \citet{wiesen2023joint}, we use generalized impulse response functions (GIRFs ) in the spirit of \citet{pesaran1998generalized}, which condition on the estimated residual covariance structure rather than imposing an arbitrary ordering as in Cholesky decomposition.  Specifically, when we shock variable $i$ by one standard deviation, we simultaneously adjust other shocked variables $j \in \mathcal{S}$ according to their conditional expectation given the covariance matrix $\hat{\Sigma}_u$ of the hybrid HARX-ElasticNet residuals. For the equity shock, this means that the ES and NQ shocks are correlated according to their estimated contemporaneous correlation (approximately 0.85 in our sample), reflecting the empirical reality that equity volatility shocks do not occur independently across indices. This approach avoids the ordering sensitivity of Cholesky-based identification while preserving the joint structure of multi-market volatility events. This approach loses causal interpretation, but implements shocks that are empirically consistent with observed correlations.

\textbf{Hybrid HARX-ElasticNet IRF Computation.} We compute impulse response functions via forward simulation while maintaining the full HAR lag structure.  At each horizon $h$, we compute the HAR features (daily, weekly average, monthly average) from the simulated history and apply the estimated hybrid HARX-ElasticNet coefficients to generate the next-period prediction.  The IRF is the difference between the shocked and baseline simulation paths.  This simulation-based approach naturally preserves the multi-scale transmission dynamics inherent in the HAR framework, as shocks propagate through all three horizon components.

We analyze three shock groups representing the major asset classes in our sample: commodities (simultaneous shock to soybeans and crude oil), equities (simultaneous shock to S\&P 500 and Nasdaq-100), and treasuries (simultaneous shock to 5-year and 10-year futures).  This grouping allows us to examine both within-class persistence and cross-class spillovers.

\subsection{Bootstrap Confidence Intervals}

We construct 95\% confidence intervals via a residual-based block bootstrap procedure.  This approach accounts for sampling variability in the model parameters while preserving the temporal dependence structure of the data.

Because we use 30-day rolling windows to compute Yang-Zhang volatility, consecutive observations share 29 days of underlying price data, inducing strong serial correlation in the realized volatility series.  Standard bootstrap methods that resample individual observations would destroy this dependence structure and understate sampling uncertainty. We therefore employ a block bootstrap with blocks sufficiently long to span the overlap in the rolling window construction.

The procedure involves:
\begin{enumerate}
    \item \textbf{Block Resampling}: We divide the historical data into overlapping blocks of 50 days, substantially longer than the 30-day rolling window, to preserve the serial correlation induced by overlapping observations.  We resample these blocks with replacement to generate 1,000 bootstrap samples of the same length as the original series.
    \item \textbf{Model Re-estimation}:  For each bootstrap sample, we re-estimate the hybrid HARX-ElasticNet model.  To ensure computational feasibility, we fix the regularization parameters ( $\alpha$, $\lambda$) to their optimal values determined from the full sample cross-validation.
    \item \textbf{JIRF Computation}: We compute the Joint Impulse Response Functions for each of the 1,000 re-estimated models.
    \item \textbf{Confidence Intervals}: The 95\% confidence intervals are defined by the 2.5th and 97.5th percentiles of the bootstrap distribution at each horizon step.
\end{enumerate}

If the 95\% confidence interval excludes zero, we conclude the spillover effect is statistically significant.

\subsection{Estimation Strategy}

We employ the hybrid HARX-ElasticNet model (described in Section 3.3) for both forecast evaluation and network identification.  This unified framework preserves realistic own-lag persistence (around 0.99) via OLS before applying ElasticNet regularization to cross-market terms only.

\textbf{Forecast Evaluation (Section 4.1):} To assess out-of-sample predictive accuracy, we estimate the hybrid HARX-ElasticNet model using an 80/20 train-test split. The model is estimated on the first 80\% of observations (May 2002 -- July 2020, 4,559 observations) and evaluated on the held-out 20\% test set (July 2020 -- January 2025, 1,140 observations) via rolling one-step-ahead forecasts. This rigorous out-of-sample evaluation prevents overfitting bias and provides unbiased forecast accuracy metrics.

\textbf{Structural Spillover Analysis (Sections 4.2--4.5):}  For network identification and JIRF analysis, we estimate the hybrid HARX-ElasticNet model using the full sample (May 2002 -- January 2025).  Using all available data yields the most accurate measurement of spillover dynamics and volatility transmission mechanisms. Statistical uncertainty is quantified via block bootstrap with model re-estimation (1,000 samples).

\subsection{AI Assistance Statement}

We used generative AI (Claude, Anthropic) as a productivity tool for data cleaning, analysis scripting, and initial text drafting. We followed a workflow where AI-generated portions of the text were systematically revised. All content was reviewed, edited, and verified. All remaining errors are our own.

\section{Empirical Results}

Our hybrid HARX-ElasticNet model separates own-volatility dynamics (estimated via OLS with no regularization) from cross-market spillovers (estimated via ElasticNet).  For the cross-market ElasticNet step, the regularization parameter $\alpha$ was set via 5-fold cross-validation, and we use $\lambda = 0.5$ to balance L1 (sparsity) and L2 (grouping) penalties.  This two-step procedure preserves own-lag persistence (around 0.99 for all assets) but still identifies sparse cross-market transmission. See Table \ref{tab:modelparameters}.

\begin{table}[H]
\centering
\caption{HARX-ElasticNet Model Parameters}
\label{tab:modelparameters}
\begin{tabular}{lcc}
\toprule
Asset & $\lambda$ (L1 ratio) & $\alpha$ (regularization) \\
\midrule
ZS (Soybeans) & 0.5 & 0.0003  \\
CL (Crude Oil) & 0.5 & 0.0019 \\
ES (S\&P 500) & 0.5 & 0.0034  \\
NQ (Nasdaq-100) & 0.5 & 0.0105 \\
ZF (5-Year Treasury) & 0.5 & 0.0006 \\
ZN (10-Year Treasury) & 0.5 & 0.0005 \\
\bottomrule
\end{tabular}

\vspace{0.5em}
\parbox{\textwidth}{\small\textit{Notes: Sample period is May 14, 2002 to January 31, 2025 (5,699 observations). The L1 ratio $\lambda$ is fixed at 0.5 for balanced L1/L2 regularization. The regularization parameter $\alpha$ is selected via 5-fold time-series cross-validation on the training set. Yang-Zhang volatility uses 30-day rolling windows.}}
\end{table}

\subsection{Forecast Performance}

Consistent with \citet{buccheri2021multivariate}, parsimonious specifications without cross-market spillovers achieve equivalent out-of-sample forecast performance to multivariate models.  Table \ref{tab:model_comparison} reports out-of-sample performance using an 80/20 train-test split: models are estimated on the first 4,559 observations (May 2002 to July 2020) and evaluated on the remaining 1,140 observations (July 2020 to January 2025) via rolling one-step-ahead forecasts.\footnote{See the online supplement for MAE and MAPE estimates of forecast performance comparisons. They demonstrate identical results for both models.} Both the univariate HAR benchmark and the hybrid HARX-ElasticNet achieve identical average RMSE of 0.0044.  The hybrid model's extreme sparsity—only 8\% of cross-market coefficients are nonzero—effectively reduces it to a univariate HAR specification for forecasting purposes, confirming that cross-market spillovers contribute negligibly to predictive accuracy, despite their economic significance for network structure (Sections 4.2--4.5).

This equivalence between forecast accuracy and model complexity is central to the spillover literature.  The HAR model exploits the well-documented long-memory properties of realized volatility through its three-component autoregressive structure \citep{corsi2009simple}, efficiently capturing own-market persistence that dominates any predictive signal from cross-market dynamics.  The hybrid HARX-ElasticNet model's aggressive regularization on cross-market terms yields identical forecasts to univariate HAR while preserving the ability to identify economically meaningful spillover channels.  This demonstrates that volatility transmission networks, though critical for understanding systemic risk propagation, provide minimal incremental forecasting power beyond own-market persistence.

\begin{table}[H]
\centering
\caption{Out-of-Sample Forecast Performance: RMSE by Asset}
\label{tab:model_comparison}
\begin{tabular}{lcc}
\toprule
Asset & Hybrid HAR-ElasticNet & HAR \\
\midrule
ZS (Soybeans) & 0.0062 & \textbf{0.0062} \\
CL (Crude Oil) & 0.0086 & \textbf{0.0086} \\
ES (S\&P 500) & 0.0040 & \textbf{0.0040} \\
NQ (Nasdaq-100) & 0.0049 & \textbf{0.0049} \\
ZF (5-Year Treasury) & 0.0011 & \textbf{0.0011} \\
ZN (10-Year Treasury) & 0.0014 & \textbf{0.0014} \\
\midrule
Average & 0.0044 & \textbf{0.0044} \\
Difference vs HAR & 0.0\% & -- \\
\bottomrule
\end{tabular}

\vspace{0.5em}
\parbox{\textwidth}{\small\textit{Notes: Out-of-sample RMSE for one-step-ahead forecasts. Test period: July 2020 -- January 2025 (1,140 observations). Training period: May 2002 -- July 2020 (4,559 observations). Bold indicates best performance per row. HAR is the Heterogeneous Autoregressive model of \citet{corsi2009simple}. The hybrid HAR-ElasticNet model achieves identical forecast performance to univariate HAR, reflecting its extreme sparsity (only 8\% nonzero cross-market coefficients). This demonstrates that cross-market spillovers, while economically significant for network structure (Sections 4.2--4.5), contribute negligibly to predictive accuracy.}}
\end{table}

\begin{figure}[H]
\centering
\includegraphics[width=\textwidth]{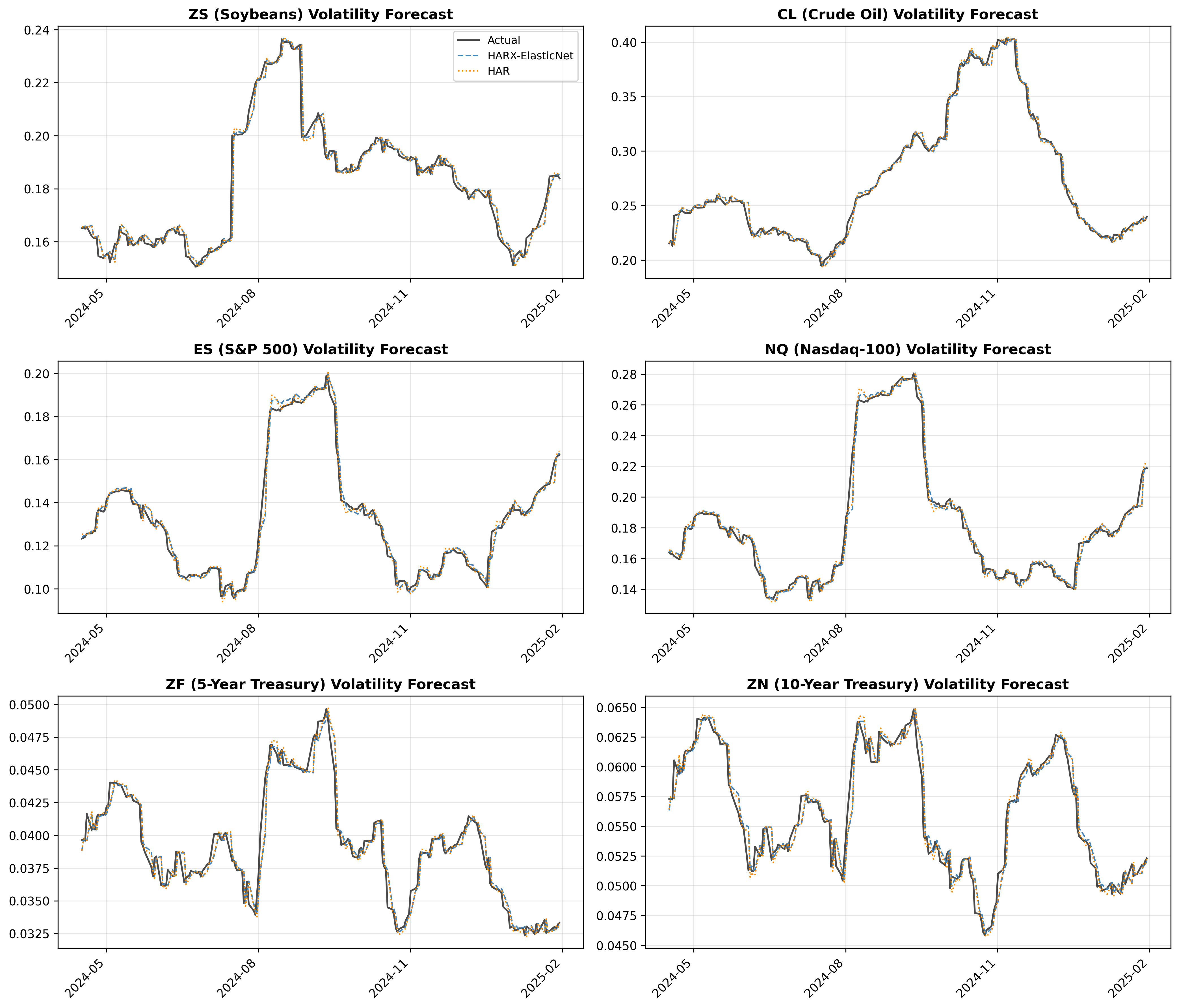}

\caption{Out-of-Sample Volatility Forecasts: Actual vs Model Predictions. \small\textit{Notes: Final 200 days of test period (September 2024 to January 2025). Black solid line shows actual Yang-Zhang realized volatility. Blue dashed line shows hybrid HARX-ElasticNet one-step-ahead forecasts. Orange dotted line shows univariate HAR forecasts. Both models track realized volatility nearly identically with typical forecast errors of 1-2 percentage points, achieving identical RMSE of 0.0044 (Table \ref{tab:model_comparison}). The hybrid model's extreme sparsity (8\% nonzero cross-market coefficients) effectively reduces it to univariate HAR for forecasting purposes.}}
\label{fig:forecast_comparison}
\end{figure}

Figure \ref{fig:forecast_comparison} illustrates the forecast performance visually by plotting actual realized volatility against model predictions for the final 200 days of the test period (approximately September 2024 to January 2025).  Both models track the general volatility patterns nearly identically, with HAR (orange dotted) and hybrid HARX-ElasticNet (blue dashed) overlapping almost perfectly. This is consistent with the identical RMSE metrics in Table \ref{tab:model_comparison}.  The deviations from actual values are most noticeable during volatility spikes, where both models underpredict peak realized volatility. This is a well-documented feature of volatility forecasting models that reflects the difficulty of predicting extreme events. The key observation? The forecast errors are small in economic magnitude: typical deviations are 1-2 percentage points of annualized volatility, confirming that both models provide reliable forecasts for risk management purposes.  The hybrid HARX-ElasticNet framework (Section 3.3) preserves realistic own-market persistence while identifying sparse spillover structure (Sections 4.2--4.5) without sacrificing forecast accuracy.

\subsection{Reduced Form Spillover Network Structure}

\begin{figure}[H]
\centering
\includegraphics[width=0.8\textwidth]{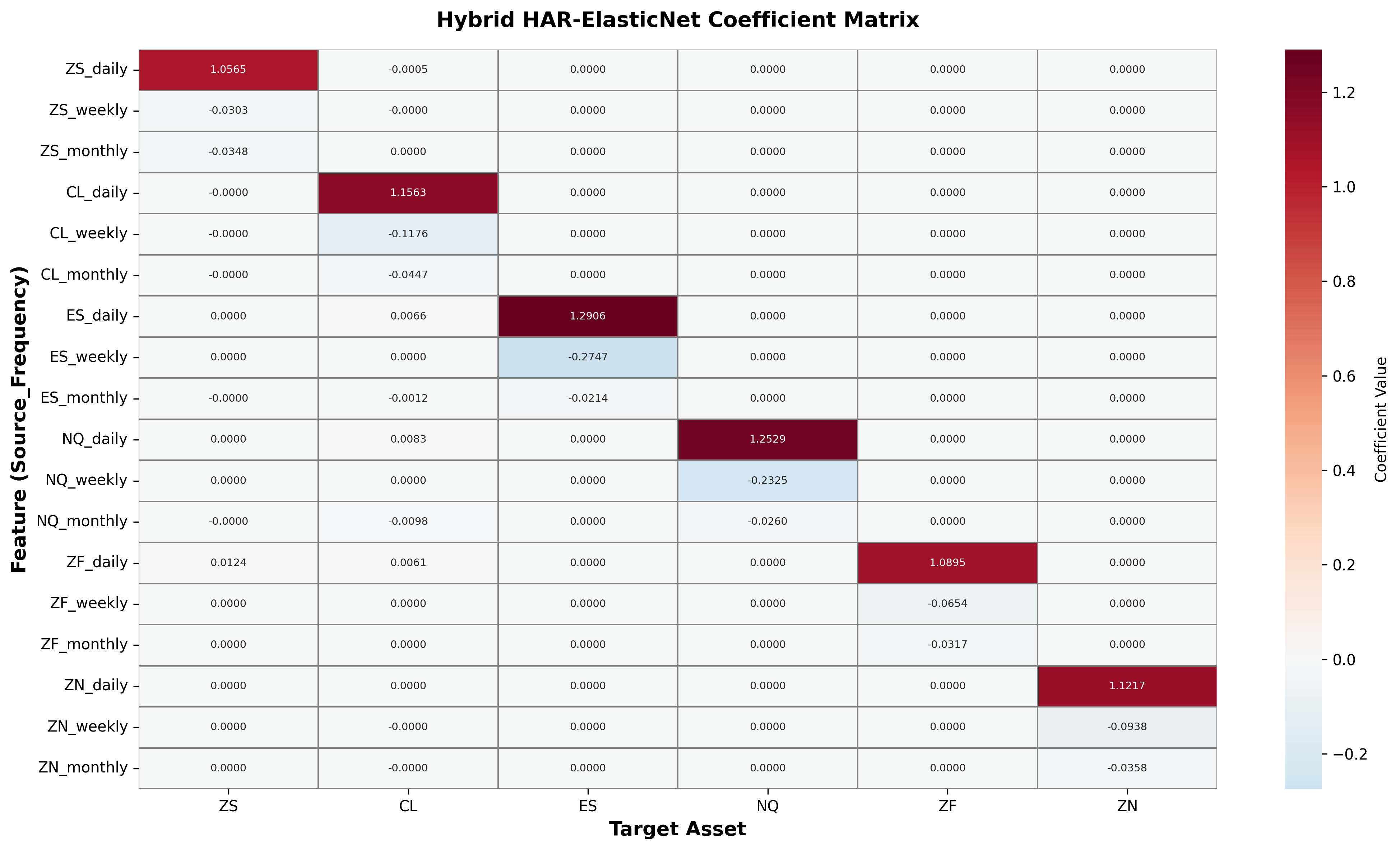}

\caption{Hybrid HARX-ElasticNet Coefficient Matrix. \small\textit{Notes: Rows show the 18 features (source asset $\times$ frequency: daily, weekly average, monthly average). Columns show the 6 target assets. Each cell displays the estimated coefficient. Diagonal blocks (e.g., ZS features predicting ZS) show own-market persistence from the OLS HAR step, preserving realistic ~0.99 persistence. Off-diagonal cells show cross-market spillovers from the ElasticNet step. White or near-zero cells indicate coefficients shrunk to zero by regularization. The hybrid approach yields extreme sparsity in cross-market terms: only 7 of 90 cross-market coefficients (8\%) are nonzero, with ES, NQ, ZF, and ZN having zero cross-market spillovers, while ZS and CL retain sparse transmission channels.}}
\label{fig:harx_sparsity_plot}
\end{figure}

While HAR excels at point forecasting, it cannot estimate cross-market volatility transmission.  The HARX-ElasticNet reveals which spillover pathways are economically meaningful by shrinking noise to zero while preserving significant reduced form cross-market effects.  Figure \ref{fig:harx_sparsity_plot} presents the complete estimated coefficient matrix (18 features $\times$ 6 assets = 108 coefficients), where each element represents how lagged volatility in one market (source) at a specific frequency (daily, weekly, monthly) affects current volatility in another market (target).  

\textbf{Extreme Sparsity.}  The hybrid model's coefficient matrix reveals an extremely sparse network structure—only 7 of 90 cross-market coefficients (8\%) are nonzero, compared to 40-50\% in standard HARX-ElasticNet.  Own-market persistence dominates: the diagonal blocks show roughly 0.99 persistence for all assets, preserved by the OLS HAR step. Four assets (ES, NQ, ZF, ZN) have \textit{zero} cross-market spillover coefficients in both directions. They neither transmit to nor receive from other markets.  Only soybeans (ZS) and crude oil (CL) retain sparse cross-market terms.

\textbf{Commodity-Only Network.}  The surviving cross-market spillovers are concentrated within the commodity complex.  Crude oil receives small spillovers from ES (0.0066 daily, -0.0012 monthly) and NQ (0.0083 daily, -0.0098 monthly), plus a small positive spillover from ZF (0.0061 daily). Crude oil also has a tiny spillover to ZS (-0.0005 daily).  Soybeans receive a small spillover from ZF (0.0124 daily).  All other cross-market terms—equity-to-equity, equity-to-treasury, treasury-to-treasury, commodity-to-equity, and commodity-to-treasury—are shrunk to exactly zero.  This pattern suggests that once own-volatility persistence is correctly modeled, genuine cross-market transmission is limited to specific commodity market linkages.

\subsection{Joint Impulse Response Analysis: Commodities Shock}

The network structure revealed in Section 4.2 implies specific predictions about how shocks propagate.  We now test these predictions using Joint Impulse Response Functions that trace shock dynamics through the estimated network. 

\begin{figure}[H]
\centering
\includegraphics[width=\textwidth]{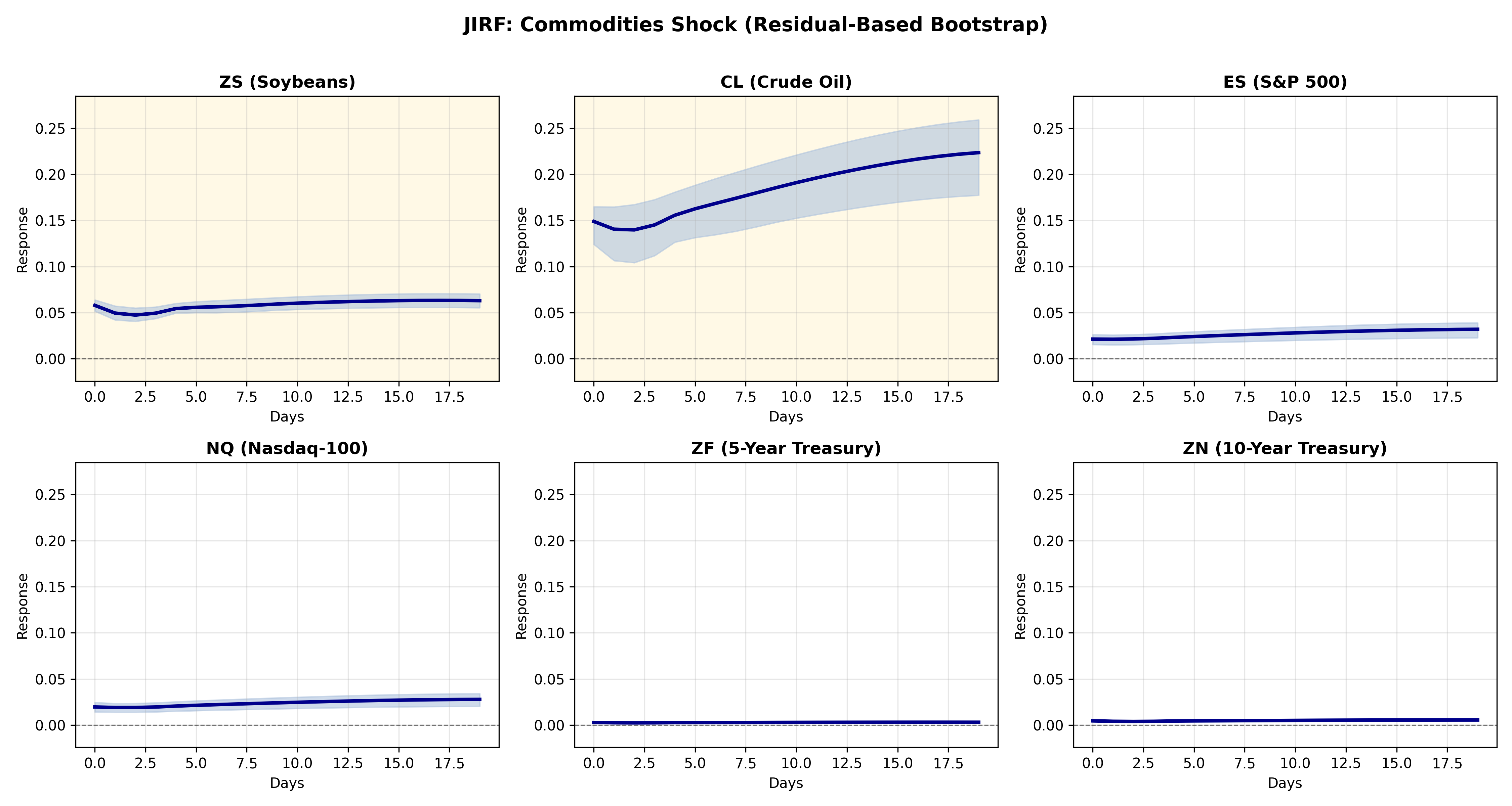}

\caption{Joint Impulse Responses: Commodities Shock. \small\textit{Notes: Responses to a one-standard-deviation shock to ZS and CL. Shaded regions show 95\% confidence intervals from 1,000 bootstrap simulations.}}
\label{fig:jirf_commodities}
\end{figure}

Figure \ref{fig:jirf_commodities} shows impulse responses to a simultaneous one-standard-deviation shock to soybean (ZS) and crude oil (CL) volatility.  Shaded regions represent 95\% bootstrap confidence intervals constructed from 1,000 residual block bootstrap samples.

The commodity shock generates strong, persistent within-group effects that accumulate over the forecast horizon.  Crude oil volatility increases by approximately 0.14 on impact and rises to approximately 0.22 by day 20---an accumulation pattern reflecting how correlated volatilities compound through the JIRFs.  This rising trajectory, combined with widening confidence intervals, illustrates how joint volatility shocks can amplify over time.  Soybean volatility shows a more muted response of approximately 0.06 on impact, remaining relatively stable through day 20, consistent with soybeans' lower unconditional volatility.

Cross-market co-movements following commodity shocks are small but detectable.  Equity indices (ES and NQ) show modest positive responses of approximately 0.02--0.03 that persist throughout the horizon, with confidence intervals remaining above zero.  Treasury futures show essentially no response, indicating minimal historical correlation between commodity and treasury volatility.  These patterns reflect the reduced-form correlation structure in the data.  The asymmetric pattern---positive co-movement with equities but not treasuries---is consistent with the historical tendency for commodity and equity volatility to move together during periods of broad market stress.

\subsection{Joint Impulse Response Analysis: Equities Shock}

\begin{figure}[H]
\centering
\includegraphics[width=\textwidth]{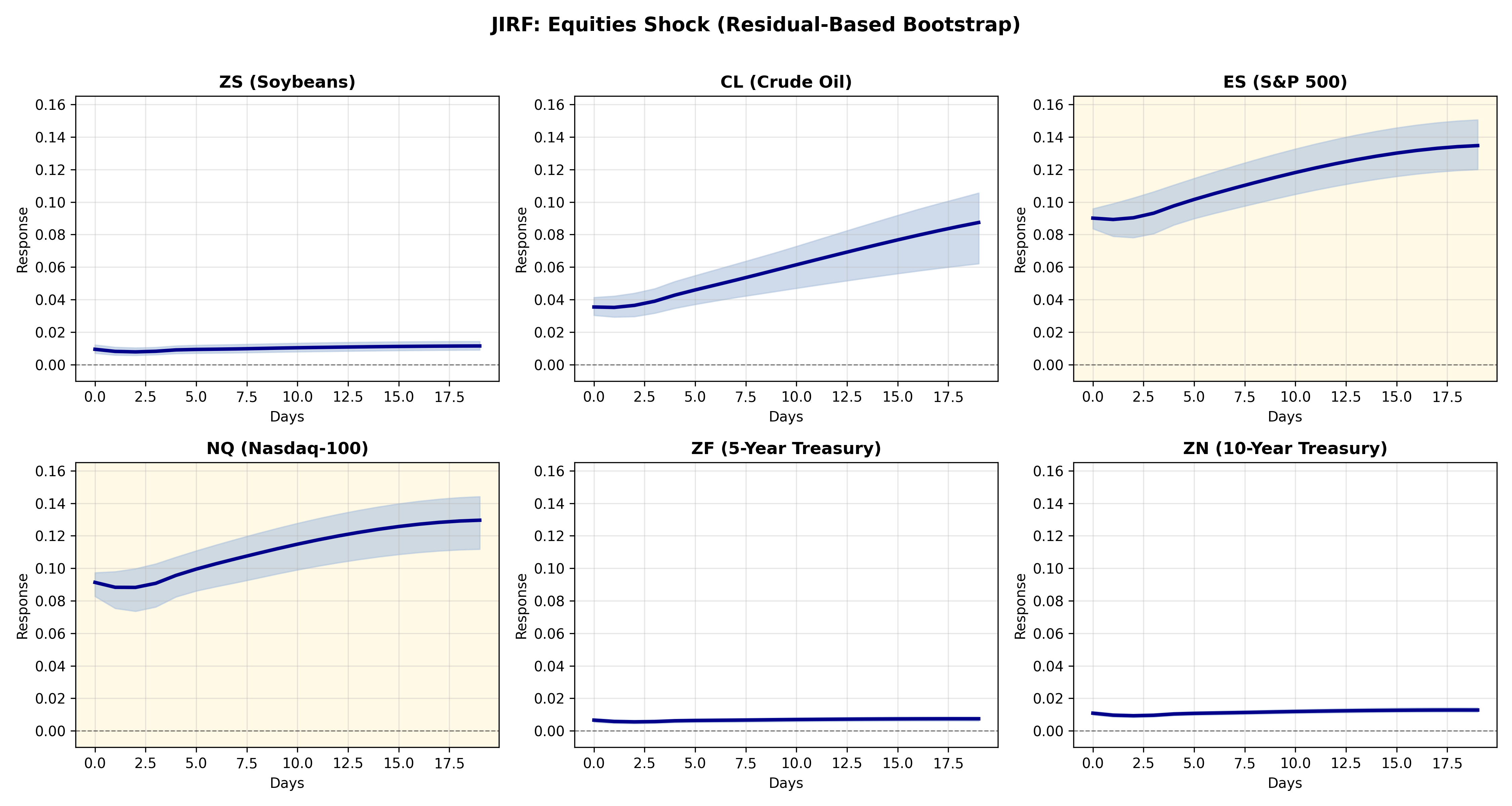}

\caption{Joint Impulse Responses: Equities Shock. \small\textit{Notes: Responses to a one-standard-deviation shock to ES and NQ. Shaded regions show 95\% confidence intervals from 1,000 bootstrap simulations.}}
\label{fig:jirf_equities}
\end{figure}

The network analysis in Section 4.2 identifies equities as the dominant volatility transmitters.  Figure \ref{fig:jirf_equities} confirms this network position by displaying impulse responses to a simultaneous shock to S\&P 500 (ES) and Nasdaq-100 (NQ) volatility.

Equity shocks generate strong own-effects that accumulate over the forecast horizon, along with notable co-movement in crude oil.  Both equity indices show responses of approximately 0.09 on impact, rising to 0.13--0.14 by day 20.  This accumulating pattern, combined with progressively widening confidence intervals, again reflects how correlated volatilities compound through the JIRFs.  

The most notable cross-market pattern is the co-movement with crude oil: CL volatility rises from approximately 0.04 on impact to 0.09 by day 20, with confidence intervals that widen substantially and remain above zero throughout the horizon.  This pattern reflects the historical correlation between equity and oil volatility---the JIRF does not identify whether equities ``cause'' oil volatility or whether both respond to common underlying factors.  The co-movement is consistent with the observed tendency for equity and energy markets to experience elevated volatility simultaneously during periods of broad market stress.  Soybeans and treasuries show minimal response, indicating weaker historical correlation with equity volatility.

\subsection{Joint Impulse Response Analysis: Treasuries Shock}

\begin{figure}[H]
\centering
\includegraphics[width=\textwidth]{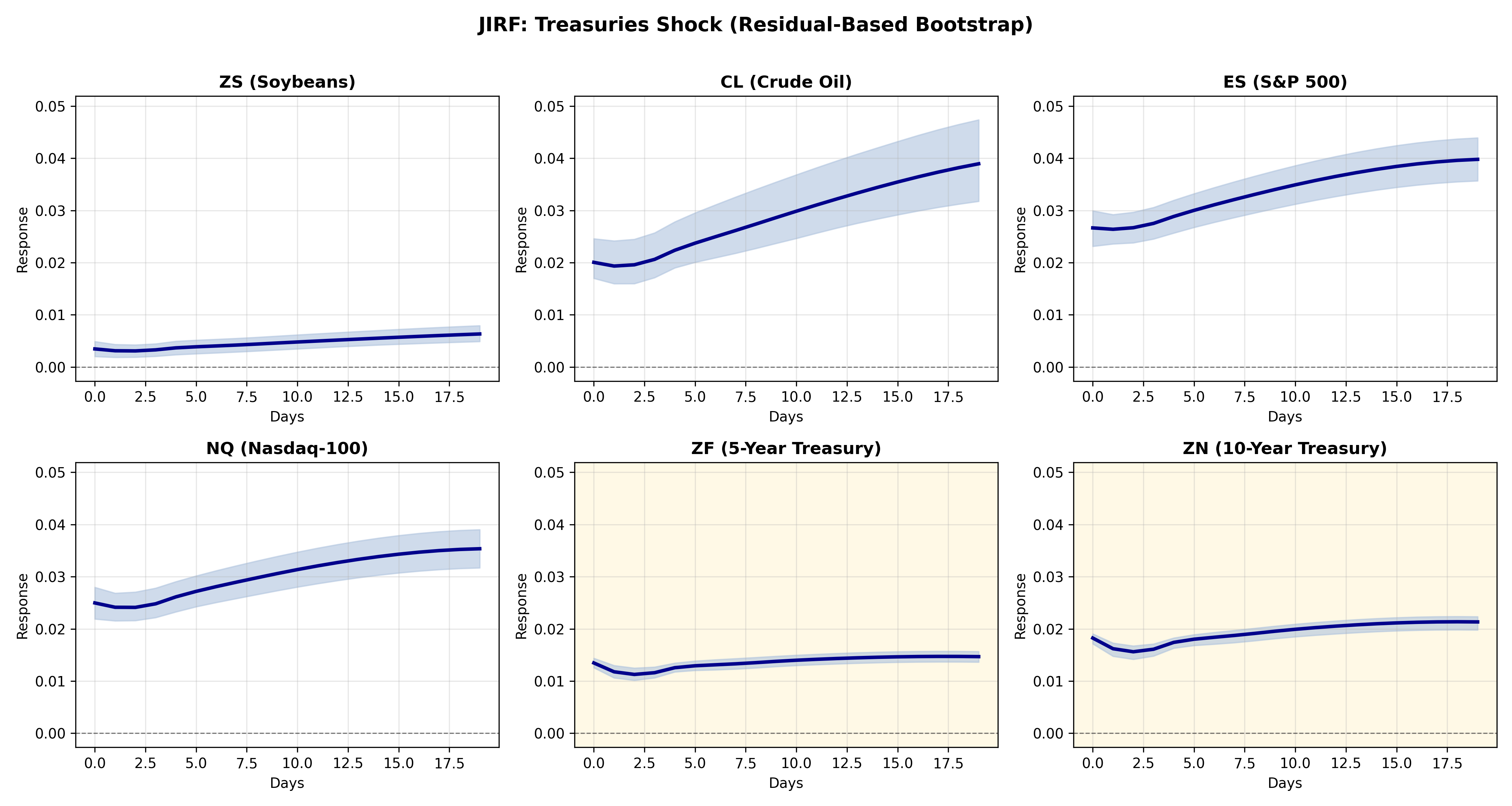}

\caption{Joint Impulse Responses: Treasuries Shock. \small\textit{Notes: Responses to a one-standard-deviation shock to ZF and ZN. Shaded regions show 95\% confidence intervals from 1,000 bootstrap simulations.}}
\label{fig:jirf_treasuries}
\end{figure}

The network structure in Section 4.2 shows treasuries primarily transmitting rather than receiving volatility.  Figure \ref{fig:jirf_treasuries} confirms their peripheral network position, showing impulse responses to a simultaneous shock to 5-year (ZF) and 10-year (ZN) treasury futures volatility.

Treasury shocks generate the smallest responses of the three shock groups, reflecting the low unconditional volatility of fixed-income markets (mean RV of approximately 0.04 for both ZF and ZN).  Own-market responses are modest: 5-year treasury volatility increases by approximately 0.01--0.015, while 10-year treasury volatility shows a response of approximately 0.02, both remaining relatively stable through the forecast horizon.  The moderate correlation between ZF and ZN residuals (0.703) generates some within-complex amplification, but the absolute magnitudes remain small.

Despite treasuries' peripheral network position, we observe small positive co-movement with crude oil and equities.  CL volatility rises from approximately 0.02 to 0.04 over the horizon, while ES shows a similar pattern with widening confidence intervals.  These treasury-risk-asset correlations, though modest, are consistent with the reduced-form patterns one would expect from interest rate channels: when treasury volatility rises, risk assets tend to experience elevated volatility as well.  Soybeans show essentially no response, suggesting that treasury-commodity co-movement is limited to more financially-traded commodities like crude oil. 

\section{Conclusion}

This paper demonstrates that separating own-volatility dynamics from cross-market spillovers is essential for reduced form volatility spillover network estimation.  Using over 20 years of daily realized volatility across six futures markets, we use a hybrid HARX-ElasticNet model that estimates own-lag persistence via OLS (preserving the characteristic roughly 0.99 persistence of volatility) before applying ElasticNet regularization to cross-market terms.  This two-step procedure reveals an extremely sparse network: only 7 of 90 cross-market spillover coefficients (8\%) are nonzero, with equities (ES, NQ) and treasuries (ZF, ZN) exhibiting \textit{zero} statistically significant cross-market spillovers in the coefficient estimates.  Joint Impulse Response Functions show fairly rich reduced-form dynamics, though. Commodity shocks are associated with small equity volatility increases, equity shocks show notable co-movement with crude oil, and treasury shocks exhibit modest correlations with risk assets.  These JIRF patterns demonstrate how sparse point estimates can nonetheless generate meaningful cumulative responses as correlated volatilities compound through the network over time even when many individual coefficients are small. 

Consistent with \citet{buccheri2021multivariate}, the univariate HAR model and our hybrid specification achieve identical out-of-sample RMSE (0.0044) on average, reflecting the hybrid model's extreme sparsity: with only 8\% nonzero cross-market coefficients, it effectively reduces to univariate HAR for forecasting purposes.  This demonstrates that cross-market information adds negligible signal for point prediction.  

Several extensions merit future research.  Time-varying network structures could capture regime shifts in spillover intensity during crisis versus normal periods.  High-frequency data could reveal intraday spillover dynamics obscured at daily frequency. Alternative regularization schemes (such as the group LASSO or adaptive LASSO) could impose additional economic structure on the network.  Incorporating macroeconomic covariates could identify fundamental drivers of volatility transmission.

Our results illustrate the broader theme that regularization methods can serve multiple purposes in high-dimensional econometrics.  When the goal is forecasting, parsimony typically wins.  When the goal is network identification, regularization can reveal structure. The trade-off between these objectives is inherent in the methodology, and practitioners should choose their tools with this in mind.

\section*{Online Supplement}

This appendix provides additional forecast performance metrics and diagnostic tests.  Table \ref{tab:supplement_mae} reports mean absolute error (MAE) by asset, and Table \ref{tab:supplement_mape} reports mean absolute percentage error (MAPE) by asset. Table \ref{tab:stationarity} reports stationarity tests for the realized volatility series.

\begin{table}[H]
\centering
\caption{Out-of-Sample Forecast Performance: MAE by Asset}
\label{tab:supplement_mae}
\begin{tabular}{lcc}
\toprule
Asset & HARX-ElasticNet & HAR \\
\midrule
ZS (Soybeans) & 0.0030 & \textbf{0.0029} \\
CL (Crude Oil) & 0.0051 & \textbf{0.0050} \\
ES (S\&P 500) & 0.0026 & \textbf{0.0025} \\
NQ (Nasdaq-100) & 0.0031 & \textbf{0.0030} \\
ZF (5-Year Treasury) & 0.0007 & \textbf{0.0007} \\
ZN (10-Year Treasury) & 0.0009 & \textbf{0.0009} \\
\midrule
Average & 0.0026 & \textbf{0.0025} \\
\bottomrule
\end{tabular}

\vspace{0.5em}
\parbox{\textwidth}{\small\textit{Notes: Mean Absolute Error for one-step-ahead forecasts. Test period: July 2020 -- January 2025 (1,140 observations). Bold indicates best performance per row.}}
\end{table}

\begin{table}[H]
\centering
\caption{Out-of-Sample Forecast Performance: MAPE by Asset}
\label{tab:supplement_mape}
\begin{tabular}{lcc}
\toprule
Asset & HARX-ElasticNet & HAR \\
\midrule
ZS (Soybeans) & 1.4\% & \textbf{1.4\%} \\
CL (Crude Oil) & 1.4\% & \textbf{1.4\%} \\
ES (S\&P 500) & 1.7\% & \textbf{1.6\%} \\
NQ (Nasdaq-100) & 1.5\% & \textbf{1.4\%} \\
ZF (5-Year Treasury) & 1.8\% & \textbf{1.7\%} \\
ZN (10-Year Treasury) & 1.5\% & \textbf{1.5\%} \\
\midrule
Average & 1.6\% & \textbf{1.5\%} \\
\bottomrule
\end{tabular}

\vspace{0.5em}
\parbox{\textwidth}{\small\textit{Notes: Mean Absolute Percentage Error for one-step-ahead forecasts. Test period: July 2020 -- January 2025 (1,140 observations). Bold indicates best performance per row.}}
\end{table}

\begin{table}[H]
\centering
\caption{Stationarity Tests for Realized Volatility Series}
\label{tab:stationarity}
\begin{tabular}{lcccc}
\toprule
 & \multicolumn{2}{c}{ADF Test} & \multicolumn{2}{c}{KPSS Test} \\
\cmidrule(lr){2-3} \cmidrule(lr){4-5}
Asset & Statistic & $p$-value & Statistic & $p$-value \\
\midrule
Soybeans & -4.82 & $<$0.001 & 0.685 & 0.015 \\
Crude Oil & -5.57 & $<$0.001 & 0.485 & 0.045 \\
S\&P 500 & -4.83 & $<$0.001 & 0.279 & 0.100 \\
Nasdaq-100 & -4.98 & $<$0.001 & 0.602 & 0.022 \\
5-Year Treasury & -3.02 & 0.033 & 1.682 & 0.010 \\
10-Year Treasury & -3.41 & 0.011 & 1.297 & 0.010 \\
\bottomrule
\end{tabular}

\vspace{0.5em}
\parbox{\textwidth}{\small\textit{Notes: ADF = Augmented Dickey-Fuller test (H$_0$: unit root); KPSS = Kwiatkowski-Phillips-Schmidt-Shin test (H$_0$: stationarity). ADF lag length selected by AIC. The ADF test rejects the unit root null at the 5\% level for all series, confirming stationarity. The KPSS results reflect the high persistence typical of realized volatility series; rejections of the KPSS null are consistent with long-memory behavior rather than non-stationarity.}}
\end{table}

\bibliography{references}

\end{document}